\documentstyle[prb,aps,twocolumn,floats]{revtex}
\input psfig.sty
\begin{document}
\draft
\title{Magnetism of a tetrahedral cluster-chain}
\author{Wolfram Brenig}
\address{Institut f\"ur Theoretische Physik, Technische Universit\"at
Braunschweig, 38106 Braunschweig, Germany}
\author{Klaus W. Becker}
\address{Institut f\"ur Theoretische Physik, Technische Universit\"at
Dresden, 01062 Dresden, Germany}
\author{Peter Lemmens}
\address{II. Physikalisches Institut, RWTH Aachen,52056 Aachen, Germany}

\date{\today}
\maketitle
\begin{abstract}
Magnetic properties of a completely frustrated tetrahedral chain are
summarized. Using exact diagonalization, and bond-operator theory results
for the ground-state phase diagram, the one-triplet excitations and the
Raman spectrum are given. The link to novel tellurate materials is clarified.
\end{abstract}

\begin{figure}[thb]
\vspace{-.7cm}
\centerline{\hspace{8cm}\psfig{file=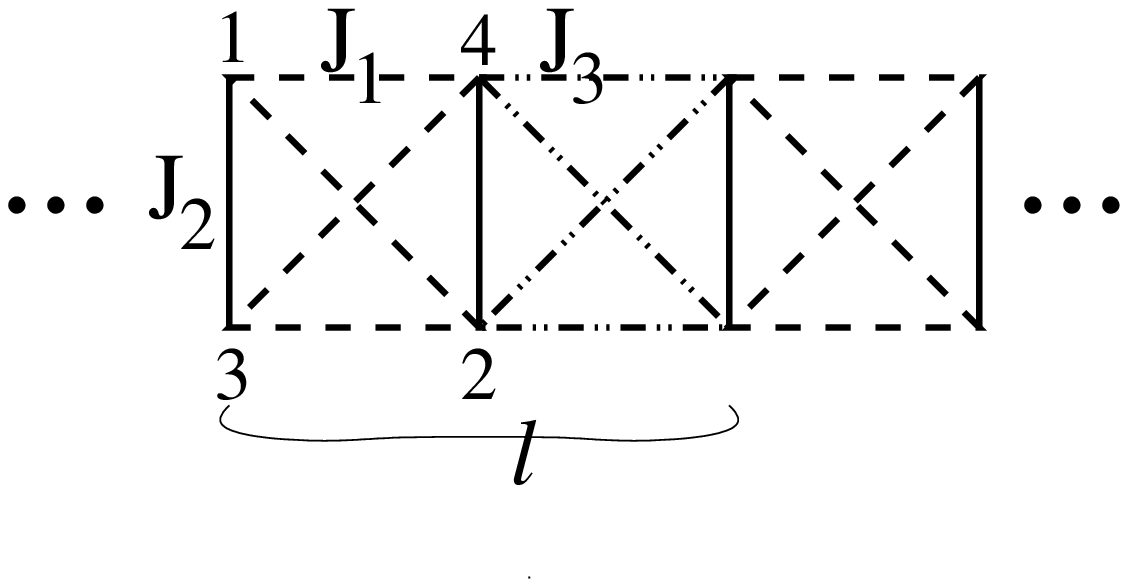,width=16cm,angle=270}}
\vspace{-7.8cm}
\caption{The tetrahedral cluster-chain. $l$ labels the unit cell
containing spin-1/2 moments at the vertices 1,...,4.} 
\label{fig3}
\end{figure}
\begin{figure}[thb]
\vspace{-.2cm}
\centerline{\psfig{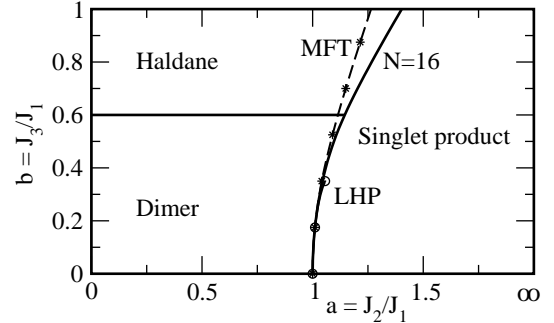}}
\caption{Phases of the tetrahedral chain.
Solid lines: exact diagonalization.
Haldane-Dimer transition at $b\approx 3/5$.
Stared (circled) dashed lines: bond-boson mean-field/MFT
(Holstein-Primakoff/LHP) approach. LHP terminates at $b=3/8$.}
\label{fig1}
\end{figure}

Recently, tellurates of type Cu$_2$Te$_2$O$_5$X$_2$ with X=Cl, Br have
been identified as a new class of spin-1/2 tetrahedral-cluster compounds
\cite{Johnsson00}. Bulk thermodynamic data has been analyzed in the limit of
isolated tetrahedra \cite{Johnsson00}. Raman spectroscopy, however
indicates substantial inter-tetrahedral coupling\cite{Lemmens01z}. In this
brief note we summarize results on the magnetism of a purely
one-dimensional (1D) chain of tetrahedra which is coupled in a geometry
analogous to that along the c-axis direction of Cu$_2$Te$_2$O$_5$X$_2$. In
this direction the exchange topology is almost completely frustrated
suggesting the spin-model of fig.
\ref{fig3}.
The hamiltonian can be written as a 1D chain in terms of the total
edge-spin operators ${\bf T}_{1(2)\,l}={\bf S}_{1(4)\,l}+{\bf S}_{3(2)\,l}$
and the dimensionless couplings $b=J_3/J_1$ and $a=J_2/J_1$
\begin{eqnarray}
\frac{H}{J_1}=
\sum_l [{\bf T}_{1l}{\bf T}_{2l}
+b {\bf T}_{2l}{\bf T}_{1l+1}
+\frac{a}{2} ({\bf T}^2_{1l} +
{\bf T}^2_{2l})  -\frac{3a}{2}] 
\label{bb1}
\end{eqnarray}
This model displays infinitely many local conservation laws: $[H,{\bf
T}^2_{i(=1,2)\,l}]=0;\,\forall\,l,i=1,2$. The Hilbert space decomposes
into sectors of fixed distributions of edge-spin eigenvalues $T_{i\,l}=1$
or $0$, each corresponding to a sequences of dimerized spin-1
chain-segments intermitted by chain-segments of localized singlets.  For
$a<a_c(b)$ the ground state is found\cite{LongVers} to be in the sector of
the infinite-length, dimerized spin-1 chain, i.e.  $T_{i\,l}=1$ for {\em
all} $i,l$, with a dimer phase at $b<b_c$ and a Haldane phase for $b>b_c$.
For $a>a_c(b)$ the infinite-length product state of singlets, i.e.
$T_{i\,l}=0$ for {\em all} $i,l$ is realized.  In the latter case the
ground state energy is $E_G=-N 3a/2$.  In the former case we have used
exact diagonalization (ED) on up to $2N=16$ sites, as well as bond-boson
theory to determine the ground state energy and phase boundaries. The phase
diagram is shown in fig.
\ref{fig1}.
Since, by ${\bf T}_{2\,l(1\,l+1)}\rightarrow {\bf T}_{1\,l(2\,l)}$,
(\ref{bb1}) is symmetric under $(J_1,a,b)\rightarrow (J_1 b,a/b,1/b)$, a
correspondingly rescaled mirror image of fig. \ref{fig1} exists.  The
combination of both covers the {\em complete} parameter space.  The
critical value of $a_c(b=0)=1$ for the 1st-order transition from the dimer
to the singlet product state agrees with \cite{Johnsson00}, while
$a^{2N=16}_c(b=1)=1.403...$ agrees with \cite{Honecker00}.
For the 2nd-order dimer--Haldane transition we find $b_c\simeq 3/5$ from
finite--size extrapolation\cite{LongVers}, which is consistent with
\cite{Kato93}.

In addition to ED, fig. \ref{fig1} displays results of an analytic
bond-boson approach to the dimerized spin-1 chain sector. Labeling the
singlet, triplet and quintet states of a tetrahedron by bosons $s,
t_\alpha, q_\alpha$, with the unit-cell index suppressed, and discarding
the high-energy quintets the edge-spins can be replaced by
$T^\alpha_{1/2}=\pm\sqrt{2/3}(t^\dagger_\alpha s + s^\dagger t_\alpha) -
i\epsilon_{\alpha\beta\gamma} t^\dagger_\beta t_\gamma/2$.  This
transforms (\ref{bb1}) into an interacting bose gas including a hardcore
constraint $s^\dagger s+ t^\dagger_\alpha t_\alpha+q^\dagger_\alpha
q_\alpha=1$. Condensing the singlets, i.e. $s^{(\dagger)}=\langle
s\rangle$, to either $\langle s\rangle=1$ (Linear Holstein Primakoff (LHP)
approximation) or to a selfconsistently determined mean  field(MFT)--value
$\langle s\rangle < 1$ the model can be diagonalized on the quadratic
level \cite{LongVers,Sachdev90a}. Contrasting the resulting ground state
energy against the singlet product state the stared(circled)-dashed phase
boundaries of fig. \ref{fig1} are obtained.  In the dimer-phase region the
agreement with ED is very good, both for LHP and MFT. In principle, the
singlet condensate restricts the bond-boson approaches to the dimer phase.
In fact, the LHP spin-gap closes at $b=3/8$ confining the LHP to
$b<3/8<b_c$. The MFT can be extended into the Haldane regime, even though
the ground-state symmetries are different, yielding a transition line
qualitatively still comparable to ED.

\begin{figure}[thb]
\vspace{-.2cm}
\centerline{\psfig{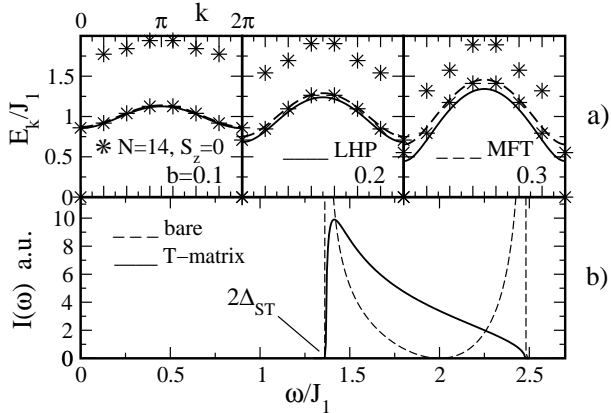}}
\caption{(a) Low lying excitations in the dimerized spin-1 chain sector.
(b) Two-magnon Raman spectrum.}
\label{fig2}
\end{figure}
Next we consider excitations in the dimer phase which would be a likely
candidate for the tellurates assuming weakly coupled tetrahedra.  The
excited states may (i) remain in the dimerized spin-1 chain sector, or
(ii) involve transitions into sectors with {\em localized} edge-singlets.
As has been pointed out in \cite{Johnsson00}, for a single tetrahedron,
the energy of a pair of two type-(ii) excitations resides in the spin-gap
of sector (i) for $1/2\leq a\leq 1$. While analogous, {\em dispersionless}
singlets are found in the spin gap of the dimer phase of the lattice
model\cite{LongVers}, we only report on one- and two-triplet excitations
of type (i) in this brief note: Figure \ref{fig2}a) displays the
dispersion of the first triplet, comparing ED, LHP, and MFT. The agreement
is very good. Figure \ref{fig2}b) displays the magnetic Raman spectrum,
i.e. total spin-zero, two-triplet excitations, based on bond-boson theory
and a Loudon-Fleury vertex with c-axis polarization. For the solid line
triplet interactions have been accounted for exactly on the two-particle
level beyond the LHP approach by a T-matrix calculation
\cite{LongVers,Kotov98}. Due to the existence of a singlet bound-state
which merges with the continuum at zero momentum the renormalized Raman
spectrum deviates strongly from the bare one.

Available Raman data on Cu$_2$Te$_2$O$_5$Br$_2$ \cite{Lemmens01z} displays
a sharp mode at 20cm$^{-1}$ and a continuum centered at 60cm$^{-1}$.  One
might speculate the sharp mode to correspond  to transitions of the
aforementioned type (ii) and the continuum to correspond to that of fig.
\ref{fig2}b). However, the measured continuum is definitely more symmetric
than the solid line in fig. \ref{fig2}b). This leaves the magnetism of the
tellurates an open issue deserving further studies.

{\bf Acknowledgements:}
This work has been presented at the 'Japanese-German bilateral
Seminar', September 2000, in Sapporo, Japan.

It is a pleasure to thank R. Valenti, C. Gros, and F. Mila
for stimulating discussions and comments. This research was
supported in part by the Deutsche Forschungsgemeinschaft
under Grant No. BR 1084/1-1.

\end{document}